\documentclass[useAMS,usenatbib]{mn2e}

\usepackage{epsfig}
\usepackage{longtable}
\usepackage{times} 
\newcommand{\sw}{$Swift$}

\def \inte {{\it INTEGRAL\,}}

\def \sw {{\it Swift}}

\def \src {\mbox{AX~J1841.0$-$0536}}

\def \igr {\mbox{IGR~J17544$-$2619}}

\def \hcm {\hbox {\ifmmode $ atom cm$^{-2}\else atom cm$^{-2}$\fi}}

\def \ATel {Astron.\ Tel.}

\def \apj {ApJ}

\def \aap {A\&A}

\def \pasj {PASJ}
\def \mnras {MNRAS}

\title[SFXT nature of AX~J1841.0$-$0536]{Confirmation of the Supergiant Fast X--ray Transient nature of \src\ from 
       \emph{Swift} outburst observations}
\author[P.\ Romano et al.]{P.\ Romano,$^{1}$ V.\ Mangano,$^{1}$ G.\ Cusumano,$^{1}$ P.\ Esposito,$^{2}$ P.A.\ Evans,$^{3}$
J.A.~Kennea,$^{4}$ 
\newauthor S.\ Vercellone,$^{1}$ V.\ La Parola,$^{1}$  H.A.\ Krimm,$^{5,6}$   D.N.~Burrows,$^{4}$  N.~Gehrels$^{6}$  \\
$^{1}$INAF, Istituto di Astrofisica Spaziale e Fisica Cosmica,
        Via U.\ La Malfa 153, I-90146 Palermo, Italy\\
$^{2}$INAF, Osservatorio Astronomico di Cagliari, localit\`a Poggio dei Pini, 
strada 54, I-09012 Capoterra, Italy\\
$^{3}$Department of Physics \& Astronomy, University of Leicester, LE1 7RH, UK\\
$^{4}$Department of Astronomy and Astrophysics, Pennsylvania State 
             University, University Park, PA 16802, USA\\
$^{5}$NASA/Goddard Space Flight Center, Greenbelt, MD 20771, USA\\
$^{6}$Universities Space Research Association, Columbia, MD, USA \\
}

\begin{document}

\date{Accepted 2010 November 29. Received 2010 November 25; in original form 2010 November 11}

\pagerange{\pageref{firstpage}--\pageref{lastpage}} \pubyear{2010}

\maketitle

\label{firstpage}

\begin{abstract}
{\it Swift} observed an outburst from the supergiant fast X--ray transient (SFXT) 
AX~J1841.0$-$0536 on 2010 June 5, and followed it with XRT for 11 days. 
The X--ray light curve shows an initial flare followed by a decay and subsequent
increase, as often seen in other SFXTs, and a dynamical range of $\sim 1600$. 
Our observations allow us to analyse the simultaneous broad-band (0.3--100\,keV) 
spectrum of this source, for the first time down to 0.3\,keV, 
which 
can be fitted well with models usually adopted to describe the emission 
from accreting neutron stars in high-mass X--ray binaries, 
and is characterized by a high absorption ($N_{\rm H}\sim 2\times10^{22}$ cm$^{-2}$), 
a flat power law ($\Gamma \sim0.2$), and a high energy cutoff.
All of these properties resemble those of the  prototype of the class, IGR~J17544$-$2619,
which underwent an outburst on 2010 March 4, whose observations we also discuss. 
We show how well AX~J1841.0$-$0536 fits in the SFXT class, 
based on its observed properties during the 2010 outburst, 
its large dynamical range in X--ray luminosity, 
the similarity of the light curve (length and shape) to those of the other SFXTs 
observed by {\it Swift}, and the X--ray broad-band spectral properties. 
\end{abstract}

\begin{keywords}
X-rays: binaries - X-rays: individual: AX~J1841.0$-$0536 - X-rays: individual: IGR~J17544$-$2619 

\noindent
Facility: {\it Swift}
\end{keywords}

%%%%%%%%%%%%%%%%%%%%%%%%%%%%%%%%%%%%%%%%%%%%%%%%

	%%%%%%%%%%%%%%%%%%%%%%%%%%%%%%%%%%%%%%%%%%%%%%%%%%%%%%%%%
	\section{Introduction\label{axj1841intro}}
	%%%%%%%%%%%%%%%%%%%%%%%%%%%%%%%%%%%%%%%%%%%%%%%%%%%%%%%%%

Supergiant fast X--ray transients (SFXTs) are a new class of High Mass
X--ray Binaries (HMXBs) discovered by \inte{} \citep[e.g.\ ][]{Sguera2005}
that are associated with OB supergiant stars via optical spectroscopy.
In the X--rays they display outbursts significantly shorter 
than those of typical Be/X--ray binaries characterized by bright flares 
with peak luminosities of 10$^{36}$--10$^{37}$~erg~s$^{-1}$ which  
last a few hours \citep[as observed by \inte; ][]{Sguera2005,Negueruela2006:ESASP604}.
As their quiescence is characterized by a luminosity of $\sim 10^{32}$~erg~s$^{-1}$ 
\citep[e.g.\ ][]{zand2005,Bozzo2010:quiesc1739n08408}, 
their dynamic range is of 3--5 orders of magnitude. 
While in outburst, their hard X--ray spectra resemble those of HMXBs 
hosting accreting neutron stars, with hard power laws below 10\,keV combined 
with high energy cut-offs at $\sim 15$--30~keV, 
sometimes strongly absorbed at soft energies \citep{Walter2006,SidoliPM2006}. 
So, even if pulse periods have only been measured for a few SFXTs,
it is tempting to assume that all SFXTs might host a neutron star. 
The mechanism producing the outbursts is still being debated,  
and it is probably related to either the properties of 
the wind from the supergiant companion 
\citep{zand2005,Walter2007,Negueruela2008,Sidoli2007} or to the 
presence of a centrifugal or magnetic barrier \citep[][]{Grebenev2007,Bozzo2008}. 

\src\ was discovered during {\it ASCA} observations of the Scutum arm 
region performed on 1994 April 12, and 1999 October 3--4 as a flaring source 
which exhibited 
flux increases by a factor of 10 (up to $\sim 10^{-10}$ erg cm$^{-2}$ s$^{-1}$)
with rising times on the order of 1\,hr  \citep{Bamba2001}, 
a strong absorption $N_{\rm H} =3\times10^{22}$ cm$^{-2}$,  
and coherent pulsations with a period of $4.7394\pm0.0008$\,s.  
A {\it Chandra} observation on 2004 May 12, which provided the coordinates 
refined to arcsecond accuracy [RA(J2000$)=18^{\rm h} 41^{\rm m} 0\fs54$,   
Dec(J2000$)=-5^{\circ}$ $35^{\prime} 46\farcs8$, \citealt{Halpern2004:18410-0535b}], 
found the source at a much fainter level ($4\times 10^{-12}$ erg cm$^{-2}$ s$^{-1}$), 
and with a spectrum that was fit with an absorbed power-law model 
[$\Gamma=1.35\pm0.30$, $N_{\rm H} =(6.1\pm1.0)\times10^{22}$ cm$^{-2}$]. 
A newly discovered source, IGR~J18410$-$0535, was observed to flare  
by \inte\ on 2004 October 8 \citep[][]{Rodriguez2004:18410-0535}, 
as it reached $\approx 70$\,mCrab in the 20--60\,keV energy range (integrated over 1700\,s)
and 20 mCrab in the 60--200\,keV range. The source was also detected in the 
20--60\,keV energy range in subsequent observations, at a flux roughly half that of 
the initial peak.  
\citet{Halpern2004:18410-0535b} identified IGR~J18410$-$0535 as \src.
\citet{Halpern2004:18410-0535a} established that the IR counterpart was 
2MASS 18410043$-$0535465, a reddened star with a weak double-peaked  H$\alpha$ 
emission line, initially classified as a Be star, which  
\citet{Nespoli2008} later reclassified as B1 Ib type star;   
this corroborated the evidence that \src\ is a member of the SFXT class,
as proposed by \citet{Negueruela2006:ESASP604}. 
\citet{Sguera2009} presented the first broad-band spectrum 
of this source, obtained with \inte\ (IBIS$+$JEM-X), that they fit with an
absorbed power-law with $\Gamma=2.5\pm0.6$, $N_{\rm H}=23^{+19}_{-14} \times 10^{22}$ cm$^{-2}$.  

In 2007 \sw\ \citep{Gehrels2004mn} observed the outburst of the periodic 
SFXT IGR~J11215$-$5952 \citep[][]{Romano2007},  which allowed us to discover that the 
accretion phase during the bright outbursts lasts much longer than a few hours, 
as seen by lower-sensitivity instruments. This is contrary to what
was initially thought at the time of the discovery of this new class 
of sources. 
Between 2007 October 26 and 2008 November 15, \src\ was observed 
by \sw\ as part of a sample of 4 SFXTs
which included IGR~J16479$-$4514, XTE~J1739--302, and \igr. 
The main aims were to characterize their long-term behavior, 
to determine the properties of their quiescent state, to monitor the onset of
the outbursts and to measure the outburst recurrence period and duration 
\citep[][]{Sidoli2008:sfxts_paperI,Romano2009:sfxts_paperV,Romano2010:sfxts_paperVI}.  
Approximately two observations per week were collected  
with the X--ray Telescope  \citep[XRT, ][]{Burrows2005:XRTmn} 
and the UV/Optical Telescope \citep[UVOT, ][]{Roming2005:UVOTmn}. 
During such an intense and sensitive monitoring, \src\ was the only SFXT 
that did not go through a bright outburst,  although several on-board Burst Alert Telescope 
\citep[BAT, ][]{Barthelmy2005:BATmn} detections have been recorded \citep[][]{Romano2009:sfxts_paperV}. 

In this paper we report on the observations of the first outburst of \src\ observed by \sw\ 
on 2010 June 5 and we compare its properties with those of the prototype of the SFXT class, 
\igr, which went into a bright outburst on 2010 March 04.

%%%%%%%%%%%%%%%%%%%%%%%%%%%%%%%%%%%%%%%%%%%%%%%%%% Figure 1 
\begin{figure}
\begin{center}
\centerline{\includegraphics[height=8.7cm,angle=-90]{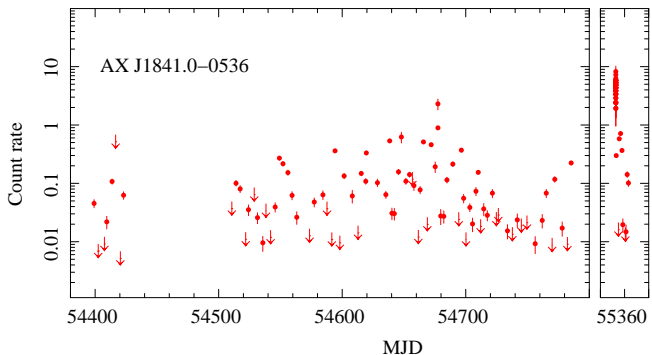}}
\end{center}
\caption{{\bf Left:} \sw/XRT (0.2--10\,keV) light curve of the 2007--2008 monitoring campaign 
\citep[2007 October 26 to 2008 November 15; ][]{Romano2009:sfxts_paperV}. 
{\bf Right:} light curve of the 2010 June 5 outburst in the same time-scale. 
The downward-pointing arrows are 3$\sigma$ upper limits.  
}
\label{axj1841fig:lcv_campaign}
\end{figure}
%%%%%%%%%%%%%%%%%%%%%%%%%%%%%%%%%%%%%%%%%%%%%%%%%%

%%%%%%%%%%%%%%%%%%%%%%%%%%%%%%%%%%%%%%%%%%%%%%%%%% Figure 2
\begin{figure}
\begin{center}
\vspace{-0.5truecm}
\centerline{\includegraphics*[angle=270,width=8.5cm]{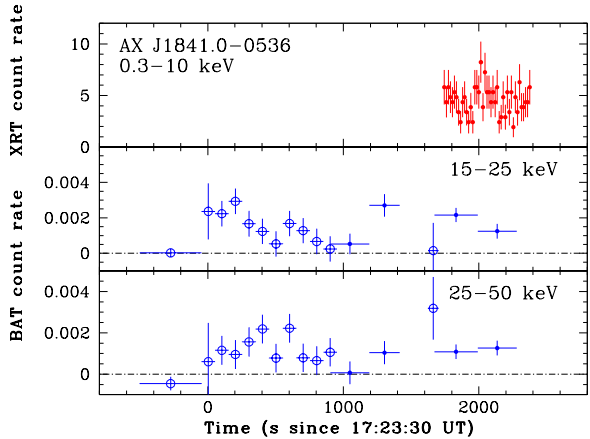}}
\end{center}
\vspace{-0.5truecm}
\caption{XRT and BAT light curves of the initial orbit of data of the 2010 June 5 outburst 
of \src\ in units of count s$^{-1}$ and count s$^{-1}$ detector$^{-1}$, respectively. 
The empty circles correspond to BAT in event mode (S/N$=5$),  
filled circles to BAT survey mode data. 
}
\label{axj1841fig:lcv_allbands}
\end{figure}
%%%%%%%%%%%%%%%%%%%%%%%%%%%%%%%%%%%%%%%%%%%%%%%%%%

 	 %%%%%%%%%%%%%%%%%%%%%%%%%%%%%%%%%%%%%%%%%%%%%%%%%%%%%%%%%%%%%%%%%%%%
 	 \section{Observations and Data Reduction\label{axj1841dataredu}}
  	 %%%%%%%%%%%%%%%%%%%%%%%%%%%%%%%%%%%%%%%%%%%%%%%%%%%%%%%%%%%%%%%%%%%%

AX~J1841.0$-$0536 triggered the \sw/BAT on 2010 June 5 at 17:23:30 UT 
\citep[trigger 423958,][]{DePasquale2010:atel2661,Romano2010:atel2662}. 
This is the first outburst of AX~J1841.0$-$0536 detected by the BAT for which 
\sw\ performed a slew, thus allowing broad-band data collection. 
The source was detected in a 1344\,s BAT image trigger, during a pre-planned observation,
and there is an indication that the source was already in outburst before this observation 
began and well after it ended. 
The XRT began observing the field rather late, at 17:51:50 UT ($T+1708$\,s), 
after the very long BAT image trigger. 
The automated target (AT, sequences 00423958000-001) observations lasted for several 
orbits, until $\sim59$\,ks after the trigger).  
Follow-up target of opportunity (ToO) observations for a total of 10.8\,ks were obtained  
(sequences 00030988093--101). 
The data cover the first 11\,d after the beginning of the outburst.  

The SFXT prototype \igr\ triggered the BAT on 
2010 March 04 at 23:13:54 UT \citep[trigger 414875,][]{Romano2010:atel2463}. 
\sw\ executed an immediate slew, so that the narrow-field instruments (NFI)
started observing it about 395\,s after the trigger. 
The AT ran for $\sim5$\,ks and was followed by one ToO observation 
(00035056149) for $\sim0.8$\,ks until the source went into Moon 
constraint.

%%%%%%%%%%%%%%%%%%%%%%%%%%%%%%%%%%%%%%%%%%%%%%%%%% Figure 3
\begin{figure}
\begin{center}
\centerline{\includegraphics[angle=270,width=7.5cm]{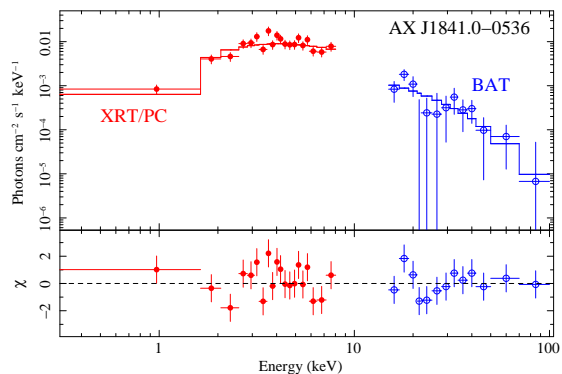}}
\end{center}
\vspace{-0.5truecm}
\caption{Spectroscopy of the 2010 June 5 outburst of \src.  
		{\bf Top:} simultaneous XRT/PC (filled red circles) and BAT (empty blue circles) data 
			fit with the {\sc highecut} model. 
		{\bf Bottom:} the residuals of the fit (in units of standard deviations). 
}
\label{axj1841:fig:meanspec}
\end{figure}
%%%%%%%%%%%%%%%%%%%%%%%%%%%%%%%%%%%%%%%%%%%%%%%%%%
%
%%%%%%%%%%%%%%%%%%%%%%%%%%%%%%%%%%%%%%%%%%%%%%%%%% Figure 4
\begin{figure}
\begin{center}
\vspace{-0.5truecm}
\centerline{\includegraphics*[angle=270,width=8.5cm]{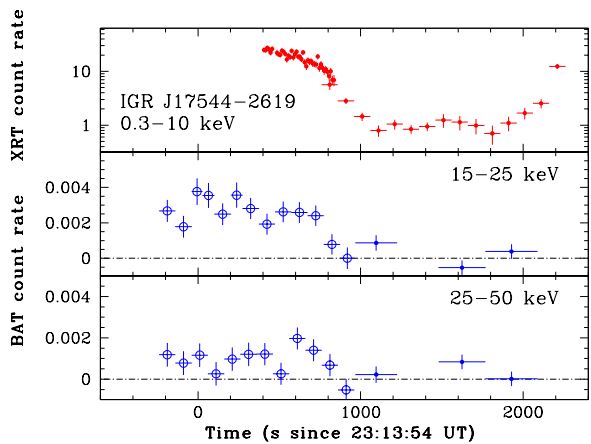}}
\end{center}
\vspace{-0.5truecm}
\caption{Same as Fig.~\ref{axj1841fig:lcv_allbands} for 
the 2010 March 4 outburst of \igr.
}
\label{axj1841fig:17544lcv_allbands}
\end{figure}
%%%%%%%%%%%%%%%%%%%%%%%%%%%%%%%%%%%%%%%%%%%%%%%%%%

The XRT data were processed with standard procedures 
({\sc xrtpipeline} v0.12.4), filtering and screening criteria by using 
{\sc ftools} in the {\sc Heasoft} package (v.6.9), as fully described in e.g.\   
\citet[][]{Romano2010:sfxts_paperVI}.   
We used the latest spectral redistribution matrices (20100930). 
The BAT data were analysed using the standard BAT analysis software within 
{\sc ftools}. Mask-tagged BAT light curves were created in several energy 
bands  \citep[see][for further details]{Romano2010:sfxts_paperVI}.  
Survey data products, in the form of Detector Plane
Histograms (DPH), are available, and were also analysed with the
standard {\sc batsurvey} software. 
All quoted uncertainties are given at 90\,\% confidence level for 
one interesting parameter unless otherwise stated.

\setcounter{table}{0} %%%%%%%%%%%%%%%%%%%%%%%%%%%%%%%%%%%%%%%%%%%%%%%%%%%%%%%%%%%%%%%%%%%%%%
\begin{table*}%%%%%%%%%%%%%%%%%%%%%%%%%%%%%%%%%%%%%%%%%%%%%   TABLE 1
 \begin{center}
 \caption{Spectral fits of XRT and BAT data of the outburst of \src\   and \igr.}
 \label{axj1841:tab:broadspec}
 \begin{tabular}{llllllrllllllll}
 \hline
 \noalign{\smallskip}
Model  &$N_{\rm H}$   &$\Gamma$           &$E_{\rm c}$    &$E_{\rm f}$    &$F$  & $L$ &$\chi^{2}_{\nu}$/dof 
  &\hspace{0.2cm}$N_{\rm H}$ &$\Gamma$    &$E_{\rm c}$    &$E_{\rm f}$    &$F$  & $L$ &$\chi^{2}_{\nu}$/dof \\
 \hline
\multicolumn{2}{l}{AX~J1841.0$-$0536} & & & & & & &  \multicolumn{2}{l}{IGR~J17544$-$2619}  \\
\sc{POW}$^{\mathrm{a}}$ &$5.2_{-2.2}^{+2.4}$ &$1.4_{-0.6}^{+0.5}$ &              &              &  $0.5$ & $2.0$ &$1.6/29$ 
      &\hspace{0.2cm} $1.9_{-0.2}^{+0.2}$   &$1.8_{-0.1}^{+0.1}$ &              &              &  $1.3$   &$2.4$ &$1.9/135$ \\ 
\sc{CPL}$^{\mathrm{a}}$ &$2.2_{-1.1}^{+1.9}$ &$0.2_{-0.6}^{+0.7}$ &$16_{-5}^{+21}$ &              &  $0.6$ & $1.8$ &$1.2/28$
      &\hspace{0.2cm} $0.7_{-0.2}^{+0.2}$   &$0.4_{-0.2}^{+0.2}$ &$7_{-1}^{+2}$  &              &  $1.6$   &$2.7$ &$1.1/134$ \\ 
\sc{HCT}$^{\mathrm{a}}$ &$1.9_{-1.0}^{+1.7}$ &$0.2_{-0.5}^{+0.4}$ &$4_{-4}^{+12}$  &$16_{-9}^{+10}$ & $0.6$ & $1.8$ &$1.2/27$ 
      &\hspace{0.2cm} $0.7_{-0.2}^{+0.2}$   &$0.6_{-0.2}^{+0.2}$ &$3_{-1}^{+1}$  & $8_{-2}^{+2}$ & $1.6$   &$2.6$ &$1.0/133$ \\  
  \hline
  \end{tabular}
  \begin{list}{}{} 
 \item{$N_{\rm H}$ is absorbing column density ($\times 10^{22}$ cm$^{-2}$); 
       $F$ is 2--10\,keV observed flux ($\times 10^{-9}$ erg cm$^{-2}$ s$^{-1}$); 
       $L$ is 2--10\,keV luminosity ($\times 10^{36}$ erg s$^{-1}$). }
  \item[$^{\mathrm{a}}$] {POW: absorbed powerlaw. CPL: cutoff powerlaw, energy cutoff E$_{\rm c}$ (keV). 
    HCT: absorbed powerlaw, high energy cutoff E$_{\rm c}$ (keV), e-folding energy E$_{\rm f}$ (keV). }
   \end{list}
  \end{center}
  \end{table*}%%%%%%%%%%%%%%%%%%%%%%%%%%%%%%%%%%%%%%%%%%%%% 

 	 %%%%%%%%%%%%%%%%%%%%%%%%%%%%%%%%%%%%%%%%%%%%%%%%%%%%%%%%%%%%%%%%%%%%
 	 \section{Results \label{axj1841results} }
  	 %%%%%%%%%%%%%%%%%%%%%%%%%%%%%%%%%%%%%%%%%%%%%%%%%%%%%%%%%%%%%%%%%%%%

 	 %%%%%%%%%%%%%%%%%%%%%%%%%%%%%%%%%%%%%%%%%%%%%%%%%%%%%%%%%%%%%%%%%%%%
 	 \subsection{Light curves \label{axj1841results:lcvs} }
  	 %%%%%%%%%%%%%%%%%%%%%%%%%%%%%%%%%%%%%%%%%%%%%%%%%%%%%%%%%%%%%%%%%%%%

Figure~\ref{axj1841fig:lcv_campaign} (left) shows the {\it Swift}/XRT 
0.2--10\,keV light curve of \src\ throughout our 2008 monitoring program
\citep[][]{Romano2009:sfxts_paperV}
background-subtracted and corrected for pile-up, PSF losses, 
and vignetting. All data in one observation (1--2\,ks, typically) were generally 
grouped as one point, except for the June 5 outburst, which shows up as a vertical line on the 
adopted scale [Fig.~\ref{axj1841fig:lcv_campaign} (right)].
The observed dynamical range of this source in the XRT band is $\approx 1600$,
considering as the lowest point a 3$\sigma$ upper limit obtained on MJD~54420 at
$5\times10^{-3}$ counts s$^{-1}$, and the highest point the peak of the June 5 outburst. 
Fig.~\ref{axj1841fig:lcv_allbands} shows the detailed light curves 
during the brightest part (first orbit) of the outburst in several energy bands. 
The BAT light curve is rather flat and weak, but significant signal is found at
the lower energies (15--50\,keV). 

For the timing analysis we converted the event arrival times to the Solar system 
barycentre with the task {\sc barycorr} and the {\it Chandra} position 
\citep{Halpern2004:18410-0535b}. 
We note that the XRT PC-mode readout frequency slightly undersamples the 
source period of $\sim$4.7 s \citep{Bamba2001} with respect 
to its Nyquist frequency, which would guarantee an unambiguous reconstruction 
of the signal. 
Timing searches were conducted in various time intervals and 
energy ranges around $\sim$4.7\,s, employing two methods: a fast-folding algorithm and an 
unbinned $Z^2_n$ test \citep{Buccheri1983}. In both cases the searches were inconclusive and we could not set 
meaningful upper limits on the pulsed fraction, because of the red noise and the scalloping 
due to the poor sampling.

	 %%%%%%%%%%%%%%%%%%%%%%%%%%%%%%%%%%%%%%%%%%%%%%%%%%%%%%%%%%%%%%%%%%%%
 	 \subsection{Spectra \label{axj1841results:spectra} }
  	 %%%%%%%%%%%%%%%%%%%%%%%%%%%%%%%%%%%%%%%%%%%%%%%%%%%%%%%%%%%%%%%%%%%%

For the 2010 June 5 outburst of \src, we extracted the mean spectrum of the 
brightest X-ray emission (obs.\ 00423958000, $T+$1708 to 2390\,s) 
and performed a fit in the 0.3--10\,keV band of the data, 
which were rebinned with a minimum of 20 counts bin$^{-1}$ to allow $\chi^2$ fitting. 
An absorbed power-law model yielded 
a column of 
$N_{\rm H}=2_{-1}^{+2}\times 10^{22}$ cm$^{-2}$, 
a photon index $\Gamma=0.6_{-0.5}^{+0.6}$ 
($\chi^2_{\nu}=1.56$, 17 degrees of freedom, dof), and 
a  2--10\,keV unabsorbed flux of  
$F_{\rm 2-10\,keV}=6.3\times10^{-10}$ erg cm$^{-2}$ s$^{-1}$. 
As there is no strict overlap between the BAT event data and the 
XRT data (see Fig.~\ref{axj1841fig:lcv_allbands}),
we extracted one spectrum from the BAT event file of the whole observation 00423958000
(`event'), and one from the XRT event file in the same observation. 
There are, however, BAT survey data available in the interval $T+1672$--$1992$\,s,
and we extracted one  (`survey') spectrum in this restricted interval. 
We performed joint fits in the 0.3--10\,keV and 14--100\,keV  
energy bands for XRT and BAT, respectively. 
A factor was included to allow for normalization uncertainties 
between the two spectra. 
The broad-band fits performed with the BAT `survey' spectrum yield consistent
results with the ones performed with the BAT `event' spectrum, albeit with 
more unconstrained parameters due to the much more limited statistics. 
Therefore in Table~\ref{axj1841:tab:broadspec} we report the fits performed with 
the BAT `event' spectrum. 
A simple absorbed power-law model is clearly an inadequate representation 
of the broad band spectrum ($\chi^2_{\nu}=1.6$ for 29 dof), so we also   
considered other models typically used to describe the X--ray emission from 
accreting pulsars in HMXBs, such as 
an absorbed  power-law model with an exponential cutoff  ({\sc cutoffpl} in {\sc xspec})
and an absorbed power-law model with a high energy cut-off ({\sc highecut}).  
The latter models provide a significantly more satisfactory fit of the 
broad-band emission, resulting in a hard powerlaw-like spectrum below 10\,keV, 
with a roll over of the higher energies when simultaneous XRT and BAT data 
fits are performed. Figure~\ref{axj1841:fig:meanspec} shows the fits
for the {\sc highecut} model. Table~\ref{axj1841:tab:broadspec} reports 
the average 2--10\,keV luminosities. 
Two estimates of the distance are available, from 
\citet[][$3.2_{-1.5}^{+2.0}$\,kpc]{Nespoli2008} and \citet[][$6.9\pm1.7$\,kpc]{Sguera2009}, 
so we assumed 5~kpc, which is consistent with both.

 	 %%%%%%%%%%%%%%%%%%%%%%%%%%%%%%%%%%%%%%%%%%%%%%%%%%%%%%%%%%%%%%%%%%%%
 	 \subsection{ \igr    \label{axj1841results:17544} }
  	 %%%%%%%%%%%%%%%%%%%%%%%%%%%%%%%%%%%%%%%%%%%%%%%%%%%%%%%%%%%%%%%%%%%%

Figure~\ref{axj1841fig:17544lcv_allbands} shows the first orbit of the 
XRT and BAT light curves of the 2010 March 04 outburst of \igr.
The XRT count rate reaches a peak exceeding 25 counts s$^{-1}$, 
then decreases to about 0.5 counts s$^{-1}$ and increases again 
up to about 20 counts s$^{-1}$ at the end of the first orbit of observations. 

\begin{figure}
\begin{center}
\centerline{\includegraphics*[angle=270,width=7.5cm]{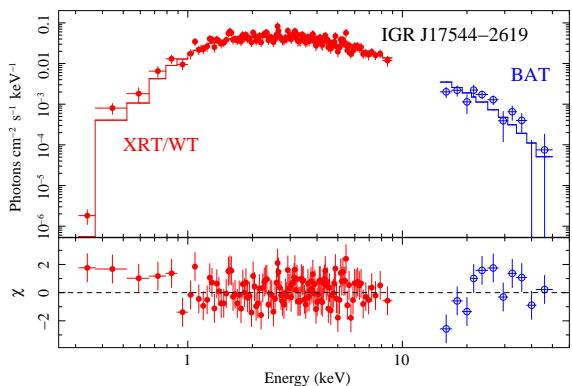}}
\end{center}
\vspace{-0.5truecm}
\caption{Same as Fig.~\ref{axj1841:fig:meanspec} for the 2010 March 4 outburst of \igr.  
}
\label{axj1841:fig:17544meanspec}
\end{figure}

The XRT/WT spectrum ($T+401$--$839$\,s), 
extracted with a grade 0 selection to mitigate residual calibration 
uncertainties at low energies, was fit with an absorbed power law 
resulting in $\Gamma=0.9_{-0.1}^{+0.1}$, 
$N_{\rm H}=(0.9_{-0.1}^{+0.2})\times 10^{22}$ cm$^{-2}$ 
($\chi^2_{\nu}=0.989$ for 125 dof), 
and $F_{\rm 2-10\,keV}\sim 1.9\times10^{-9}$ erg cm$^{-2}$ s$^{-1}$ (unabsorbed). 
The XRT/PC spectrum ($T+840$ to $T+2242$\,s) yields
$N_{\rm H}=(1.8_{-0.5}^{+0.7})\times 10^{22}$ cm$^{-2}$,  $\Gamma=1.5_{-0.4}^{+0.4}$ ($\chi^2_{\nu}=748$, 17 dof) 
and $F_{\rm 2-10\,keV}\sim 1.9\times10^{-10}$ erg cm$^{-2}$ s$^{-1}$ (unabsorbed). 
We extracted a BAT spectrum strictly simultaneous with the XRT one 
and fitted them (0.3--10\,keV, 14--50\,keV) 
with the same models as adopted for \src. 
The results are in Table~\ref{axj1841:tab:broadspec}, while   
Figure~\ref{axj1841:fig:17544meanspec} shows the fits
for the {\sc highecut} model. 
For the luminosity calculation we adopted a distance of 
3.6~kpc  \citep{Rahoui2008}.

%%%%%%%%%%%%%%%%%%%%%%%%%%%%%%%%%%%%%%%%%%%%%%%%%% Figure 5
%%%%%%%%%%%%%%%%%%%%%%%%%%%%%%%%%%%%%%%%%%%%%%%%%%

 	 %%%%%%%%%%%%%%%%%%%%%%%%%%%%%%%%%%%%%%%%%%%%%%%%%%%%%%%%%%%%%%%%%%%%
 	 \section{Discussion \label{axj1841discussion}}
  	 %%%%%%%%%%%%%%%%%%%%%%%%%%%%%%%%%%%%%%%%%%%%%%%%%%%%%%%%%%%%%%%%%%%%

In this paper we report our analysis of the 2010 June 5 outburst of \src\ and
the 2010 March 04 outburst of the SFXT prototype \igr. 
While in the first case, the image trigger was a very long one and NFI data 
could be collected only $\sim 1700$\,s after the trigger, when the source was 
relatively dim, in the second case, the slew occurred immediately after the trigger,
while \igr\ was still very bright.  

\begin{figure}
\begin{center}
\vspace{-0.5truecm}
\centerline{\includegraphics[width=8.5cm,angle=0]{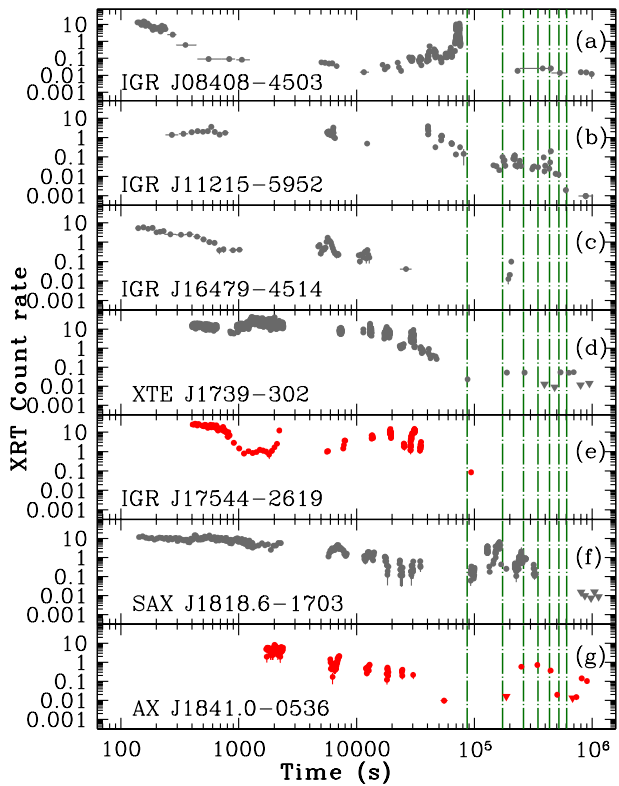}}
\end{center}
\vspace{-1truecm}
\caption[XRT light curves]{Light curves of the most representative outbursts of SFXTs 
followed by {\it Swift}/XRT referred to their respective BAT triggers
(IGR~J11215$-$5952 did not trigger the BAT, so it is referred to MJD 54139.94).
Points denote detections, triangles 3$\sigma$ upper limits.  
Red data points (panels e, g) refer to observations presented here for the first time,
while grey points to data presented elsewhere. 
Where no data are plotted, no \sw\ data were collected.  
Vertical dashed lines mark time intervals equal to 1 day, up to a week. 
References: IGR~J08408--4503 \citep[2008-07-05, ][panel a]{Romano2009:sfxts_paper08408};
IGR~J11215$-$5952 \citep[2007-02-09, ][panel b]{Romano2007}; 
IGR~J16479$-$4514 \citep[2005-08-30, ][panel c]{Sidoli2008:sfxts_paperI}; 
XTE~J1739$-$302 \citep[2008-08-13, ][panel d]{Sidoli2009:sfxts_paperIV}; 
SAX~J1818.6$-$1703 \citep[2009-05-06, ][panel f]{Sidoli2009:sfxts_sax1818}. 
Panels e and g report the 2010-03-04 outburst of IGR~J17544$-$2619 
and the  2010-06-05 outburst of AX~J1841.0$-$0536, respectively (this work).
}
\label{axj1841fig:best_sfxts}
\end{figure}

Figure~\ref{axj1841fig:best_sfxts} (panels e and g) shows the full light 
curves of the outbursts of \src\ and \igr\ as they were observed by \sw\ 
for 11 and 2 days after the trigger, respectively. 

The \src\ XRT light curve shows a decreasing trend from the initial bright flare 
from a maximum of $\sim 8$ counts s$^{-1}$ down to $\sim 0.01$ counts s$^{-1}$ during the
first day, with several flares superimposed, hence yielding a dynamic range of
approximately 900 during this outburst.  
Then, after three days, the source count rate rose again and reached $\sim 1$ counts s$^{-1}$. 
We estimate that the observed dynamical range of this source in the XRT 
band, considering the historical data we collected during our monitoring campaign 
\citep[][see Fig.~\ref{axj1841fig:lcv_campaign}]{Sidoli2008:sfxts_paperI,Romano2009:sfxts_paperV} 
is $\approx 1600$, hence  placing it well in the customary range for SFXTs. 

The outburst of \igr\ has similar characteristics to the one observed on 2008 March 31, 
as the XRT light curve shows a peak at about 25 counts s$^{-1}$, 
decreases to about 0.5 counts s$^{-1}$ and then increases again 
up to about 20 counts s$^{-1}$ at the end of the first orbit  
(Fig.~\ref{axj1841fig:17544lcv_allbands}). 
This behaviour was previously observed in \igr\ and, most notably,  
in IGR~J08408$-$4503 \citep{Romano2009:sfxts_paper08408} and 
SAX~J1818.6$-$1703 \citep{Sidoli2009:sfxts_sax1818}, so that this multiple-peak
structure of the light curve could be considered a defining characteristic of the 
SFXT class and it is likely due to inhomogeneities within the accretion flow \citep[e.g.][]{zand2005}.
 
Figure~\ref{axj1841fig:best_sfxts} compares the light curves of \src\ and \igr\ 
with the outbursts of SFXTs as observed during our monitoring campaigns with \sw. 
The most complete set of X--ray observations of an outburst of a SFXT is 
the one of the periodic SFXT IGR~J11215$-$5952 
\citep[][]{Romano2007,Sidoli2007,Romano2009:11215_2008}, which 
was surprisingly long. 
We now know that such a length of the outburst (hence the length of the accetion 
phase) is a common characteristic of the 
whole sample of SFXTs followed by \sw, and in this respect \src\ 
fits right in, as its outburst lasted several days.

%%%%%%%%%%%%%%%%%%%%%%%%%%%%%%%%%%%%%%%%%%%%%%%%%% Figure 6

We have presented the broad-band (0.3--100\,keV) simultaneous 
spectroscopy of \src. This allows us to make a comparison with the findings on
the other SFXTs that were observed in the same fashion. 
The soft X--ray spectral properties observed during this flare are 
generally consistent with those observed with {\it ASCA} during the 1999 flare 
\citep[][$N_{\rm H} =3\times10^{22}$ cm$^{-2}$, $\Gamma=1$]{Bamba2001}.
As \src\ was observed relatively late after the trigger, no meaningful information 
can be derived on variability of the soft spectral parameters during the outburst, 
such as the absorbing column density. 
However, we note that the value of $\Gamma$ in outburst follows the same trend of 
`harder when brighter' as reported in table~4 of \citet[][]{Romano2009:sfxts_paperV},
which was based on out-of-outburst emission.

For the joint BAT$+$XRT spectrum during the 2010 June 5 outburst, 
an absorbed power-law model is an inadequate description, 
and more curvy models are required. We considered 
an absorbed power-law model with an exponential cutoff and 
an absorbed power-law model with a high energy cut-off,
models typically used to describe the X--ray emission from 
accreting neutron stars in HMXBs.  
We obtained a good fit of the 0.3--100\,keV spectrum, characterized by 
high absorption $N_{\rm H} \sim 2\times10^{22}$ cm$^{-2}$, 
a hard power law below 10\,keV, and a high energy cutoff.
These properties of \src\ are reminiscent of those of the prototypes of the SFXT 
class, \igr\ [whose data we have presented here and in 
\citet{Sidoli2009:sfxts_paperIII,Sidoli2009:sfxts_paperIV,Romano2010:sfxts_paperVI}],
and XTE~J1739$-$302 (\citealt{Sidoli2009:sfxts_paperIII,Sidoli2009:sfxts_paperIV}).

Although no statistically significant pulsations were found in the 
present data, \src\ is one of the 4 SFXTs with known  pulse period 
\citep{Bamba2001}, $P_{\rm spin} =4.7394\pm0.0008$\,s, the others being 
IGR~J11215$-$5952 \citep[186.78$\pm$0.3\,s, ][]{Swank2007:atel999},
IGR~J16465$-$4507  \citep[228$\pm$6\,s, ][]{Lutovinov2005}, and    
IGR~J18483$-$0311  \citep[21.0526$\pm$0.0005\,s, ][]{Sguera2007}. 
While lacking the detection of cyclotron lines, which 
would yield a direct measurement of the magnetic field $B$ of the 
neutron star, an indirect estimate can be obtained by 
considering the {\sc highecut} fit to the broad-band spectrum of \src\
in outburst. Our value of the high energy 
cutoff $E_{\rm c} < 16$\,keV, although loosely constrained, 
yields a $B\la 3\times$10$^{12}$~G \citep{Coburn2002}. 
This value for $B$, which is indeed similar  
to the one derived for the prototype of the SFXT class \igr,  
is inconsistent with a magnetar nature of \src. 

In conclusion, we have shown how AX~J1841.0$-$0536 nicely fits in the SFXT class, 
based on the observed properties of \src\ during the 2010 June 5 outburst:  
a large dynamical range in X--ray luminosity, 
the similarity of the light curve length and shape to those of the prototype of the class \igr, 
and the X--ray broad-band spectrum, which we show here for the first time 
down to 0.3\,keV, thus constraining both the absorption and the cutoff energy.

  \vspace{-0.7cm}
 
%%%%%%%%%%%%%%%%%%%%%%%%%%%%%%%%%%%%%%%%%%%%%%%%%%%%%%%%%
\section*{Acknowledgments}
%%%%%%%%%%%%%%%%%%%%%%%%%%%%%%%%%%%%%%%%%%%%%%%%%%%%%%%%%

We thank the \sw\ team duty scientists and science planners 
and the remainder of the \sw\ XRT and BAT teams,
S.\ Barthelmy in particular, for their invaluable help and support. 
This work was supported in Italy by contract ASI-INAF I/009/10/0, 
at PSU by NASA contract NAS5-00136. 
PE acknowledges financial support from the Autonomous Region of Sardinia 
through a research grant under the program PO Sardegna FSE 2007--2013, L.R. 7/2007 
``Promoting scientific research and innovation technology in Sardinia''.

\setcounter{table}{1} %%%%%%%%%%%%%%%%%%%%%%%%%%%%%%%%%%%%%%%%%%%%%%%%%%%%%%%%%%%%%%%%%%%%%%
 \begin{table*}
 \begin{center}
 \caption{Summary of the {\it Swift} observations.\label{axj1841:tab:obs} }
 \begin{tabular}{lllllll}
% \hline
 \hline
 \noalign{\smallskip}
Source & Sequence   & Obs/Mode  & Start time  (UT)  & End time   (UT) & Exposure & Time since trigger   \\ 
     &      &           & (yyyy-mm-dd hh:mm:ss)  & (yyyy-mm-dd hh:mm:ss)  &(s)  & (s)       \\
  \noalign{\smallskip}
 \hline
 \noalign{\smallskip} 
AX~J1841.0$-$0536 &00423958000	&BAT/evt   &2010-06-05 17:15:13         &2010-06-05 19:01:13     &1515 & -502\\
                  &00423958000	&XRT/PC    &2010-06-05 17:52:02	&2010-06-05 18:03:24	 &682	  &1708    \\
                  &00423958001	&XRT/PC    &2010-06-05 19:02:59	&2010-06-06 09:46:49	 &6111    &5965    \\
                  &00030988093	&XRT/PC    &2010-06-07 21:07:22	&2010-06-07 21:23:58	 &969	  &186228  \\
                  &00030988094	&XRT/PC    &2010-06-08 14:49:26	&2010-06-08 15:05:56	 &977	  &249952  \\
                  &00030988095	&XRT/PC    &2010-06-09 16:18:40	&2010-06-09 16:38:56	 &1212    &341705  \\
                  &00030988096	&XRT/PC    &2010-06-10 19:47:57	&2010-06-10 21:33:56	 &1414    &440663  \\
                  &00030988097	&XRT/PC    &2010-06-11 13:10:31	&2010-06-11 13:35:56	 &1503    &503217  \\
                  &00030988098	&XRT/PC    &2010-06-13 13:41:14	&2010-06-13 15:23:58	 &848	  &677859  \\
                  &00030988099	&XRT/PC    &2010-06-14 04:09:44	&2010-06-14 05:58:57	 &1554    &729969  \\
                  &00030988100	&XRT/PC    &2010-06-15 02:39:58	&2010-06-15 04:22:57	 &1184    &810983  \\
                  &00030988101	&XRT/PC    &2010-06-16 03:55:52	&2010-06-16 04:14:58	 &1110    &901937  \\

IGR~J17544$-$2619 &00414875000  &BAT/evt &2010-03-04 23:10:00    &2010-03-04 23:30:02    &1202    & -239 \\  
                  &00414875000	&XRT/WT  &2010-03-04 23:20:40	 &2010-03-05 00:47:11	  &421     &401     \\
                  &00414875000	&XRT/PC  &2010-03-04 23:27:59	 &2010-03-05 00:49:12	  &1520    &840     \\
                  &00414875001	&XRT/WT  &2010-03-05 01:21:58	 &2010-03-05 08:48:39	  &783     &7679    \\
                  &00414875001	&XRT/PC  &2010-03-05 01:22:03	 &2010-03-05 09:02:56	  &2502    &7684    \\
                  &00035056149	&XRT/PC  &2010-03-06 00:59:19	 &2010-03-06 01:33:50	  &801     &92721   \\
  \noalign{\smallskip} 
  \hline
  \end{tabular}
  \end{center}
  \end{table*}
%%%%%%%%%%%%%%%%%%%%%%%%%%%%%%%%%%%%%%%%%%%%%%%%%%%%%%%%%%%%%%%%%%%%%%%%%%%%%%

\bsp

\label{lastpage}

\end{document}